Pomazanov M.V.[1]

# Second-order accuracy metrics for scoring models and their practical use
(revised 18.11.2022)

## Abstract

The paper proposes new second-order accuracy metrics for scoring/rating models, which show the target preference of the model - it is better to diagnose "good" objects or better to diagnose "bad" ones for a constant generally accepted predictive power determined by the first-order metric - the Gini index. There are two metrics, they have both an integral representation and a numerical one. The numerical representation of metrics is of two types, the first of which is based on binary events to evaluate the model, the second on the default probability given by the model. Comparison of the results of calculating the metrics allows you to validate the calibration settings of the scoring/rating model and reveals its distortions. The article provides examples of calculating second-order accuracy metrics for ratings of several rating agencies, as well as for the well-known approach to calibration based on van der Burg's ROC curves.

[1]National Research University Higher School of Economics, Russian Federation,
  Email: mhubble@yandex.ru,
  ORCID: 0000-0003-2377-0131,
  URL: https://www.hse.ru/en/org/persons/500361

# 1 Introduction and problem statement

The most popular method for validating scoring (rating) models currently used in practice is the analysis of the characteristics of the Cumulative Accuracy Profile (CAP) profile and its integral statistical accelerator - the accuracy coefficient or Gini index. A detailed explanation of this method can be found in [12]. To construct a CAP curve, all obtained credit scoring / rating values are sorted in ascending order for the entire set of scoring objects (for example, legal entities or individuals) for a certain period or for a set of periods. It should be known whether a default occurred or not for an object during the analyzed period (usually a year) from the date of the rating calculation. The rating model should have predictive power and separate future defaulters from "good" targets by placing defaulters at the top of the list.

For each possible numerical value of the rating (say, from 0 to 100), for example, 10, the Y-axis is the proportion of defaults with a lower rating than 10, the X-axis is the proportion of all objects with a lower rating than 10. An effective rating model The CAP curve should be as convex as possible above the diagonal and ascending at the start of the X-axis. If the CAP curve is close to the diagonal, it means that the rating does not in any way characterize the probability of default (ineffective scoring model, Random-model).

The Gini index or Accuracy Ratio (AR) of a scoring model is the ratio of the area between the CAP curve and the diagonal to the area between the ideal CAP curve and the diagonal, (see Fig. 1), $AR = \frac{a_R}{a_P}$. Value of AR is measured in the range from 0 (ineffective system) to 1 (ideal system), if AR is less than zero (maybe up to -1), then the rating model is worse than its absence, or the rating objects are ordered in reverse order.

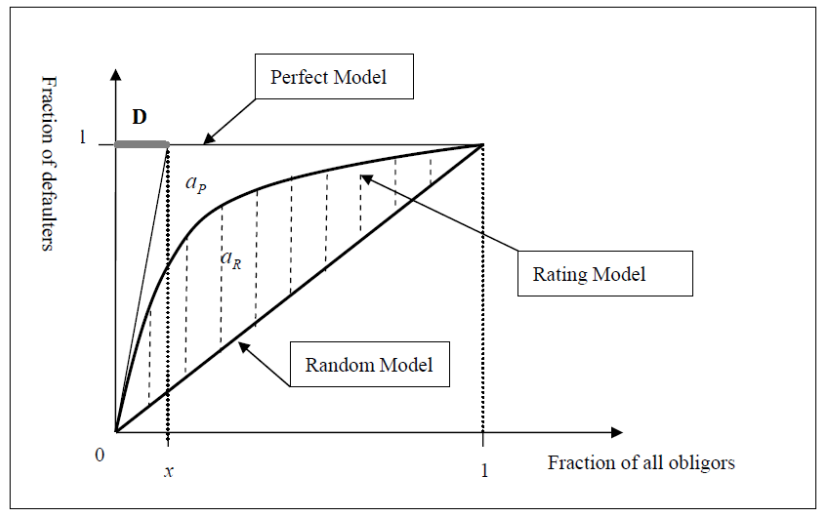

*Figure   1: CAP curve.*

The ideal CAP-curve will be set in the form of a triangle with a notch with the base D indicated in Fig. 1, where D is the share of defaults among all rated objects of the training set.

Exact expressions for AR

$$AR = \frac{2\int_0^1 C(x)dx - 1}{1 - D} \qquad (1)$$

where C(x) is the CAP-curve value for the x-coordinate of the scoring object, which denotes the proportion of objects sorted by rating, so that the rating of the object x is higher or equal to the x proportion of all rated objects, but lower than the opposite proportion of 1-x, $C(0) = 0$, $C(1) = 1$. From the construction of the CAP-curve it follows that the probability of default (PD) in the training sample of objects x∈[0,1] is given by the product

$$PD(x) = D \cdot \frac{d}{dx} C(x) \qquad (2)$$

To plot the Receiver Operating Characteristic (ROC) curve, the OX axis represents the share of non-default objects sorted by rating (scoring) ascending order, so there is no triangle with base D in the ROC curve figure. The ideal ROC curve will coincide with the OY axis up to the value of one, therefore, the dependence on the share of defaults D will formally disappear from formula (1) for the Gini index

$$AR = 2 \cdot AUC - 1, \qquad (3)$$

where $AUC = \int_0^1 R(g)dg$ is area under ROC curve.

The AUC metric has long been widely used in various problems of diagnosing binary events [4]: from the problems of signal-noise separation in radio engineering [7] and medical diagnostics [13], [16] to algorithms artificial intelligence [3] and credit scoring used in risk management [8].

The Gini exponent calculated from the ROC curve turns out to be invariant with respect to D. Indeed, the share of "good" objects (i.e, no default) will be defined as

$$g(x) = \frac{x - \int_0^x PD(s)ds}{1-D}, \qquad (4)$$

where, obviously, $g(x) \in [0,1]$ and

$$C(x) = R(g(x)), \quad R(0) = 0, \quad R(1) = 1. \qquad (5)$$

Then $AR = 2\int_0^1 R(g(x))dg(x) - 1 = 2\frac{\int_0^1 C(x)(1-PD(x))}{1-D}dx - 1 = \frac{2\int_0^1 C(x)dx - 1}{1-D}$, that is, formulas (1) and (3) are equivalent.

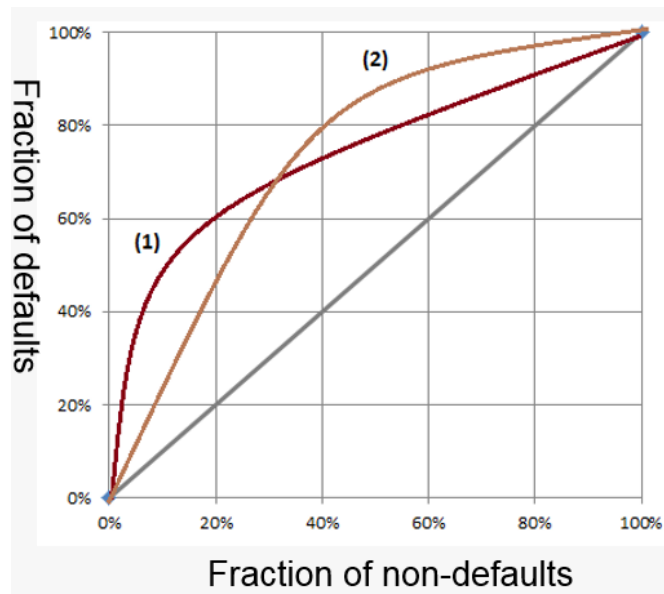

Figure 2: ROC curves of two models of the same power (Gini).

In Fig. 2 shows the ROC curves of two scoring models having the same AR accuracy. However,

curve type 1 characterizes the rating model 1, which is better than model 2 at identifying "bad" objects - this is called the left-hand preference of the model. On the contrary, curve type 2 shows that model 2 is better at separating "good" objects - this is called right-hand preference. Mathematically, this is easy to verify if you pay attention to the geometric meaning of formula (2), which determines the probability of default as the tangent of the slope of the tangent for the CAP curve, multiplied by the normalizing default.

The Gini metric is the most thoroughly studied in credit risk management, in the work [5] estimates of the statistical accuracy of this metric on limited measurement data are given. However, the Gini metric does not characterize the peculiarities of the preference of scoring models in the best way to determine the "bad" ones and not to distinguish the "good" objects in the best way, or vice versa.

In practice, taking into account the target preference of the scoring / rating model is in demand, including its purposeful construction for a given preference. For example, if the lending strategy is aimed at the maximum level of approval when lending to approved companies on general terms (collateral, margin, short loan term, etc.), then it is reasonable that such a "credit pipeline" would be preferred to model 1. On the contrary, in the case of long-term lending on concessional terms or lending in a downturn, a reasonable preference will be given to model 2, because such a strategy requires the selection of the best companies with a minimum level of risk.

To separate these goals, it is necessary to introduce additional metrics that can distinguish the features of the scoring model with left-hand and right-hand preference for the convexity of the ROC curve.

## 2  Definition of metrics of the second order of accuracy of scoring models

### 2.1  Definition

It is assumed that a scoring (rating) model has been built for which a convex continuous ROC curve is known $y = R(x)$, the inverse function is defined $x = R^{-1}(y)$.

**Definition**

1. *The Left-hand AR of the scoring model is the LAR metric*

$$LAR = 2 \cdot LAUC - 1, \quad LAUC = \int_0^1 \frac{\int_0^c R(x)dx}{c \cdot R(c)} dc;$$

2. *The Right-hand AR of the scoring model is called the RAR metric*

$$RAR = 2 \cdot RAUC - 1, \quad RAUC = \int_0^1 \frac{\int_f^1 (1-R^{-1}(y))dy}{(1-f) \cdot (1-R^{-1}(f))} df.$$

By changing variables, it is easy to see that the equivalent expression for RAUC is

$$RAUC = \int_0^1 \frac{\int_c^1 (R(x) - R(c))dx}{(1-c) \cdot (1-R(c))} \cdot R'(c) dc$$ or, similar to RAR, the expression

$$RAR = 1 - 2 \cdot \int_0^1 (1 - R(c)) \cdot \left( \int_0^c \frac{R'(x)}{(1-x) \cdot (1-R(x))} dx \right) dc.$$

With AR = 0 (random scoring model) LAR = RAR = 0. Indeed, at zero AR, $R(x) = x$ and $LAR[random] = 2\int_0^1 \frac{\int_0^c x dx}{c^2} dc - 1 = 0$, the same will happen for RAR.

The right-hand RAR metric is a representation of the left-hand LAR metric and vice versa, if you do the symmetric inverse transformation of the ROC curve. Namely:

$$\begin{cases} y = 1 - R(x) \\ Q = 1 - x \end{cases} \quad (6)$$

Then Q(y) will be a right-hand ROC curve (curve type 2 Fig. 2) if R(x) was a left-hand ROC curve (curve type 1 Fig. 2) and vice versa.

## 2.2 AR, LAR, RAR metrics on a finite set of scoring objects

Let the statistical binary data on the credit risk indicator be provided, including:

1. $N$ – non-defaults, with ratings $S_n$, $n = 1 \ldots N$

2. $D$ – defaults with ratings before default $\hat{S}_d$, $d = 1 \ldots D$.

3. all ratings are ordered from «bad» to «good» $S_{n+1} \geq S_n$, $\hat{S}_{d+1} \geq \hat{S}_d$.

The default rate in a given sample will obviously be $PD = \frac{D}{D+N}$. Let the function

$$\delta_u(w) = \begin{cases} 1, & \text{if } u > w \\ \frac{1}{2}, & \text{if } u = w \\ 0, & \text{if } u < w \end{cases}.$$

ROC curve $R_n$ at point $x_n = 0 \ldots \frac{n}{N} \ldots 1$ calculated as

$$R_0 = 0, R_n = \frac{1}{D} \cdot \sum_{d=1}^{D} \delta_{S_n}(\hat{S}_d) \quad (7)$$

The first-order discriminating metric AR (Gini coefficient) is calculated using the well-known formula of the Mann-Whitney statistics:

$$AUC = \frac{1}{N \cdot D} \cdot \sum_{n=1}^{N} \sum_{d=1}^{D} \delta_{S_n}(\hat{S}_d), \quad AR = 2 \cdot AUC - 1 \quad (8)$$

The second-order left-hand and right-hand metric LAR and RAR is calculated by the formulas:

$$LAUC = \frac{1}{N} \sum_{k=1}^{N} \begin{vmatrix} \frac{\sum_{n=1}^{k} \sum_{d=1}^{D} \delta_{S_n}(\hat{S}_d)}{k \cdot \sum_{d=1}^{D} \delta_{S_k}(\hat{S}_d)} & \text{if } \sum_{d=1}^{D} \delta_{S_k}(\hat{S}_d) \neq 0 \\ 0 & \text{if } \sum_{d=1}^{D} \delta_{S_k}(\hat{S}_d) = 0 \end{vmatrix},$$

$$RAUC = \frac{1}{D} \sum_{d=1}^{D} \begin{vmatrix} \frac{\sum_{n=d}^{D} \sum_{k=1}^{N} \delta_{S_k}(\hat{S}_n)}{(D-d+1) \cdot \sum_{k=1}^{N} \delta_{S_k}(\hat{S}_d)} & \text{if } \sum_{k=1}^{N} \delta_{S_k}(\hat{S}_d) \neq 0 \\ 0 & \text{if } \sum_{k=1}^{N} \delta_{S_k}(\hat{S}_d) = 0 \end{vmatrix}, \quad (9)$$

$$LAR = 2 \cdot LAUC - 1, \quad RAR = 2 \cdot RAUC - 1$$

.

Formulas (9) are obtained from the definition of second-order left-hand and right-hand metrics after replacing the integral with finite differences taking into account (7). It should only be noted that the computational volume of calculating metrics is of the order of magnitude $D \cdot N$ for AR, $D \cdot N^2/2$ for LAR, $N \cdot D^2/2$ for RAR. Therefore, in the case of big data, you will have to resort to the random thinning technique to save computational resources.

The statistical error of the LAR, RAR metrics at the time of publication of this work has not been investigated, however, the statistics of the AR indicator has been thoroughly studied [5]. It is proposed to use these results to formulate approximate interval criteria taking into account the finiteness of measurements. However, given the absence of formulas for LAR, RAR statistics, there is no point in "chasing accuracy," so we use conservative estimates of the AR standard deviation. In [1], an upper bound for the standard deviation AR for a convex continuous ROC curve depending on the AR value is presented in the form

$$\sigma_{AR} = \sqrt{\frac{(2N+1) \cdot (1-AR^2) - (N-D) \cdot (1-AR)^2}{3 \cdot N \cdot D}}. \tag{10}$$

Which is in practical terms of the study of the scoring model at $N \gg D \gg 1$

can be simplified to $\sigma_{AR} \cong \sqrt{\frac{1+2AR-3AR^2}{3 \cdot D}}$.

## 2.3 AR, LAR, RAR metrics imputed to the calibrated rating model

Let the number of non-default rating segments $S_i$, ordered in descending order of PD dimensions, is equal to N, $i = 1 \ldots N$. For each segment of the rating $S_i$ of the test sample containing $n_i \geq 0$ objects, the calibrated probability of default is determined $d_i = PD(S_i)$.

Assuming $g_0 = 0$, $R_0 = 0$, coordinates $g_k = \frac{\sum_{i=1}^{k} n_i \cdot (1-d_i)}{\sum_{i=1}^{N} n_i \cdot (1-d_i)}$, $R_k = \frac{\sum_{i=1}^{k} d_i \cdot n_i}{\sum_{i=1}^{N} d_i \cdot n_i}$ will form the imputed ROC-curve in the axes OX, OY, respectively.

The dotted numerical analog of the definitions of the Gini AR, as well as the LAR and RAR metrics on the imputed ROC curve, will be the following expressions:- the imputed Gini is calculated as

$$AUC_{imp} = \sum_{k=1}^{N} \left[ \frac{R_k + R_{k-1}}{2} \cdot (g_k - g_{k-1}) \right], AR_{imp} = 2 \cdot AUC_{imp} - 1. \tag{11}$$

- imputed second-order accuracy metrics LAR, RAR are calculated as

$$LAUC_{imp} = \sum_{k=1}^{N} \begin{bmatrix} \frac{g_k - g_{k-1}}{g_k \cdot R_k} \cdot \sum_{s=1}^{k} \frac{R_s + R_{s-1}}{2} \cdot (g_s - g_{s-1}), & \text{if } g_k \cdot R_k \neq 0 \\ 0, & \text{if } g_k \cdot R_k = 0 \end{bmatrix},$$

$$RAUC_{imp} = \sum_{k=1}^{N} \begin{bmatrix} \frac{R_k - R_{k-1}}{(1-g_{k-1}) \cdot (1-R_{k-1})} \cdot \sum_{s=k}^{N} \left(1 - \frac{g_s + g_{s-1}}{2}\right) \cdot (R_s - R_{s-1}), \\ \text{if } (1 - g_{k-1}) \cdot (1 - R_{k-1}) \neq 0 \\ 0, \quad \text{if } (1 - g_{k-1}) \cdot (1 - R_{k-1}) = 0 \end{bmatrix}, \tag{12}$$

$$LAR_{imp} = 2 \cdot LAUC_{imp} - 1, \quad RAR_{imp} = 2 \cdot RAUC_{imp} - 1.$$

For a specific rating model, the calculations of the first and second order metrics are possible using formulas (8),(9) at the development-validation stage according to the default data of the test sample objects. Calculations according to formulas (11),(12) are carried out after the rating model is calibrated and the objects are filled with rating categories for the current non-default sample. It is of practical interest to compare the results (8),(9) vs (11),(12), because it lie in base of validation of the calibration.

## 3. Limit convex ROC curves

The simplest geometric illustration of a ROC curve with asymmetric scoring model properties is a triangular ROC curve (see Fig. 3).

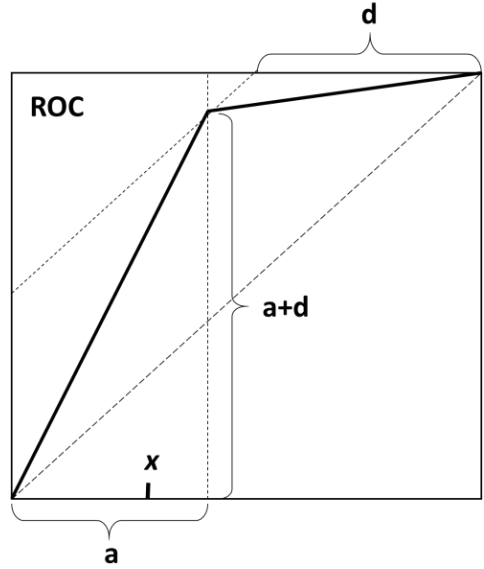

Figure 3: Triangular ROC curve.

This ROC curve is set with only two parameters $a, d$ at that $a \in [0, 1-d]$. Obviously, the Gini index is $AR = d$. To find LAR and RAR, it is necessary to represent the ROC-curve as a function

$$ROC(x) = \begin{cases} x\frac{a+d}{a}, & x \leq a \\ x\frac{1-a-d}{1-a} + \frac{d}{1-a}, & x > a, \end{cases} \quad (13)$$

then use the formulas LAR and RAR from the definition of paragraph 2.1.

Omitting rather cumbersome but standard calculations, we get

$$LAR(a, d) = a \cdot \ln(a) - \frac{1-a}{1-a-d} \cdot (a+d) \cdot \ln(a+d)$$
$$RAR(a, d) = (1-a-d) \cdot \ln(1-a-d) - \frac{a+d}{a} \cdot (1-a) \cdot \ln(1-a) \quad (14)$$

From formulas (14), the limiting values of the metrics LAR and RAR for a given accuracy of the AR scoring model follow. If the value $a$ on the right tends to zero ($a \to 0 + \epsilon$), then we get the ROC-curve, shifted to the maximum to the left, with the maximum LAR and minimum RAR. If $a \to 1 - d - \epsilon$, it will turn out the other way around, i.e. minimum LAR and maximum RAR. It is not difficult to calculate the limits (14)

$$\min(LAR, RAR) = AR + (1-AR) \cdot \ln(1-AR), \quad \max(LAR, RAR) = -\frac{AR \cdot \ln(AR)}{1-AR}. \quad (15)$$

Formula (15) gives the range of possible LAR, RAR values for a convex ROC curve
$$LAR, RAR \in [\min(LAR, RAR), \max(LAR, RAR)].$$

# 4. Trapezoidal representation of the ROC curve, segment of indifference

Triangular metrics $a(LAR)$ and $a(RAR)$ obtained from a given ROC curve with a known Gini index ($AR = d$) have an obvious geometric meaning. If we solve the transcendental equations (13), then they will parameterize the ROC-curve of the given scoring model, indicating the size and extent of its three main zones of default probability. You can conditionally divide these zones according to the traffic light principle: to the «red» - $(0, a(LAR)]$ , to «yellow» - $(a(LAR), a(RAR)]$ and on "green" - $(a(RAR), 1]$ along the ROC axis «Fraction of non-defaults» (see Fig. 5). The color depth of the "traffic light" coloring will be determined by the slope of the triangular ROC (13), which is a multiple of the expected probability of default PD of the entire population[2]. In the «red» zone, PD has a multiplier $PD_{red} = PD \cdot \mu_L$, similarly in the «green» zone is $PD_{green} = PD/\mu_R$ and in the "yellow" zone there is, on average, a neutral position $PD_{yellow} = PD$.

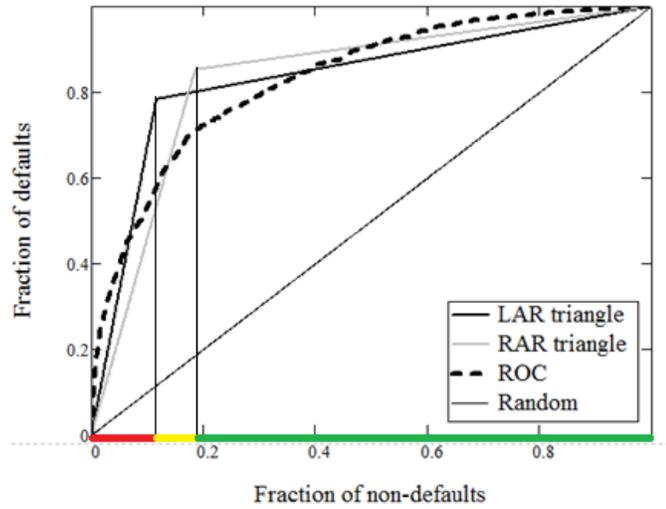

*Figure 4. Triangulation of the ROC-curve of the scoring model AR = 0.667, LAR = 0.53, RAR = 0.486 (a(LAR)=0.116, a(RAR)=0.185)*

The multipliers $\mu_L$ , $\mu_R$ of the left and right zones of the ROC curve (among the population of non-default objects) are calculated as

$$\mu_L(LAR, AR) = \frac{a(LAR)+AR}{a(LAR)}$$

$$\mu_R(RAR, AR) = \frac{1-a(RAR)}{1-a(RAR)-AR}$$
(16)

and $\mu_L > 1, \mu_R > 1$.

The higher the multiplier $\mu L$, the higher the left preference for the ROC curve (L-type), on the other hand, the higher $\mu R$, the higher the right target preference at equal AR (R-type). With an increase in the overall discriminatory accuracy AR, both indicators will increase monotonically. From formulas (14), (16) will follow a unified equation for the multipliers

$$sAR = \frac{AR}{\mu_s-1} \ln\left(\frac{AR}{\mu_s-1}\right) - \frac{\mu_s-1-AR}{\mu_s(1-AR)-1} \cdot \frac{AR \cdot \mu_s}{\mu_s-1} \ln\left(\frac{AR \cdot \mu_s}{\mu_s-1}\right),$$
(17)

---

[2] This is not a strict definition, but it becomes precise as $PD \to 0$.

where $s = L, R$. Thus, the solution of equation (17) is the function $\mu = \mu_s(sAR, AR)$, which has the single solution in the area $AR \in (0,1)$,

$sAR \in \left(AR + (1 - AR) \cdot \ln(1 - AR), -\frac{AR \cdot \ln(AR)}{1 - AR}\right)$,

wherein $a(LAR) = \frac{AR}{\mu_L - 1}$, $a(RAR) = \frac{\mu_R - 1 - \mu_R \cdot AR}{\mu_R - 1}$.

The parameters $AR, \mu_L, \mu_R$ (17) can also be an alternative parameterization of the ROC curve instead of $AR, LAR, RAR$.

Using the multipliers $\mu_L, \mu_R$ you can get the ranges of the trapezoidal representation of the ROC-curve Fig.5, which has two properties:

1. The area under the trapezoid coincides with the AUC of ROC curve;
2. The upper base of the trapezoid is parallel to the lower. That is, in the non-default sampling range of the OX axis $I \in (a, 1 - b)$, the trapezoidal ROC curve is indifferent to default discrimination.

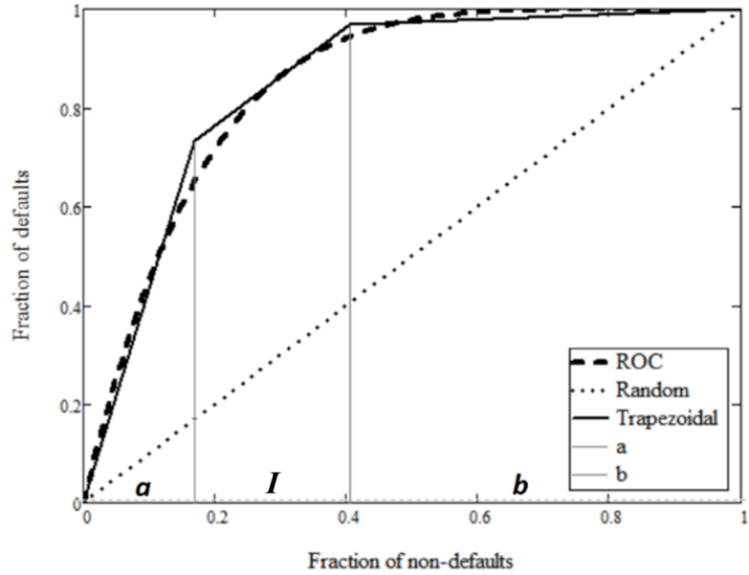

*Figure 5. Trapezoidal transformation of a ROC curve*

Analytical calculations give

$\boldsymbol{a} = \frac{a(LAR)}{1+\sqrt{1-A}}$, $\boldsymbol{b} = \frac{1-a(RAR)}{1+\sqrt{1-A}}$,

where $A = AR \cdot \left(1 + \frac{1}{\mu_L - 1} + \frac{1}{\mu_R - 1}\right)$.

Whence it follows that the length of the neutral section is

$\boldsymbol{I = 1 - a - b = \sqrt{1 - A}}$. (18)

The coordinates of the two points of the upper base of the trapezoidal ROC curve are the lower $(a; a \cdot \mu_L)$ and upper $(1 - b; 1 - \frac{b}{\mu_R})$, respectively.

The length of the neutral section $I$ has a simple meaning from the point of view of the binary choice model. Namely, (18) is the concentration (percentage) of binary choice objects that have quasi-neutrality (or indifference) to discrimination. For example, if we consider the ROC curve as a population ranking curve by income, then metric (18) will estimate the concentration of the population with an average income (middle-class). At the same time, metric (18) is not invariant with respect to ROC preference (L-type, where there are few rich people or R-type, where there are many poor people). Moreover, under condition $A = 1$ (equivalent to $\frac{1}{\mu_L - 1} + \frac{1}{\mu_R - 1} = \frac{1 - AR}{AR}$), the middle-class as such disappears, and the ROC-curve takes on a triangular shape Fig. 3, indicating only two income classes (like to patricians and plebeians in ancient Rome).

## 5. Practical samples of rating/scoring models preference

One of the most popular PD calibration models for scoring cards is the logit form
$$PD = \frac{1}{1 + e^{\alpha \cdot R + \beta}},$$
where $\alpha, \beta$ – constant parameters, $R$ – scoring score. In order to answer the question whether this form of calibration is suitable for different target preferences of scoring / rating models, let us consider a similar form of calibrations, which follows from van der Burgt's one-parameter class of ROC curves [15]. Van der Burgt's ROC curve model type is easy to analyze and recommended for calibrating scoring models [14]. The one-parameter family is
$$ROC(x) = \frac{1 - e^{-kx}}{1 - e^{-k}}.$$

Gini index (AR) is calculated as a function
$$AR(k) = 2 \cdot \left( \frac{1}{1 - e^{-k}} - \frac{1}{k} - \frac{1}{2} \right), \tag{19}$$
for which there is an inverse one $k(AR)$.

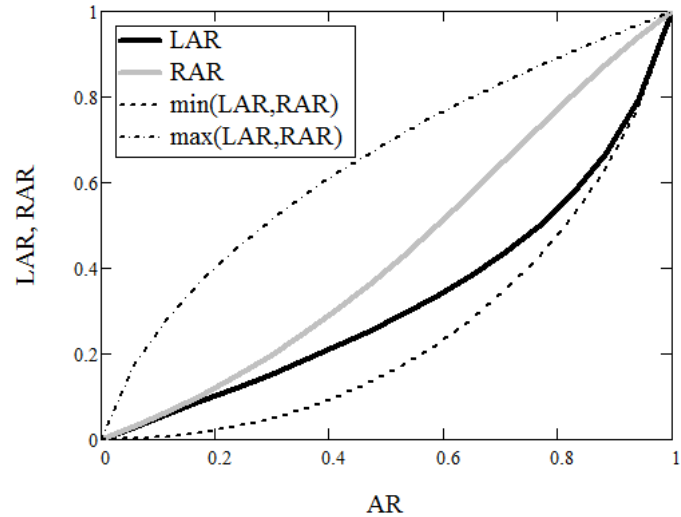

*Figure 6. Dependence of the left-hand metric LAR and the right-hand one - RAR on the Gini index AR for the van der Burgt's curves family's.*

Using the definition from paragraph 2.1 and (19), for the van der Burgt's curve, $LAR(k)$, $RAR(k)$, $k(AR)$ is calculated and $LAR(AR)$, $RAR(AR)$ dependencies are plotted (see Fig. 6). The dependences min(LAR, RAR) and max(LAR, RAR) on AR are established by relations (15).
From the graphs Fig. 6 it follows that the van der Burgt's curve has an exclusively right-hand target preference, since RAR>LAR always fulfilled on the range of the Gini index AR ∈ (0,1).

Therefore, the van der Burgt's curve cannot be used to calibrate the left-hand target preference scoring models.

It would be interesting to know what target preferences are among well-known rating agencies? Rating agencies do not use any calibration function for some implied rating score, but set the rating category of the rated object through strictly regulated manipulations with the assessments of risk-dominant parameters. Then, for each rating grade, they make a historical estimate of the default rate, which is updated annually using historical data.

Metrics for assessing the accuracy of discrimination of the first and second order for rating agencies can be estimated directly by binary events (default / non-default on the annual horizon) of rated objects using formulas (8),(9), but this requires access to their closed databases. However, there is a way to evaluate these same metrics from imputed data using the approach (11),(12). To do this, you need to extract only two entities - this is the distribution of rated objects (number or proportion) by rating categories (or classes) and the calibration values of PD in each rating category.

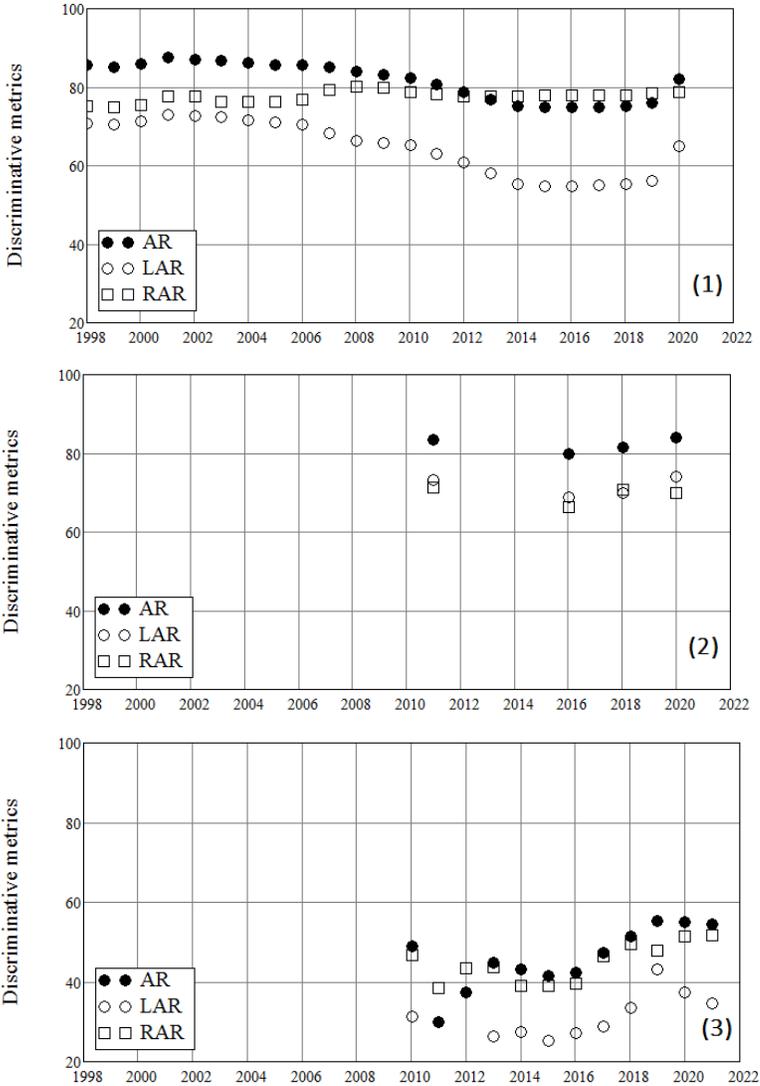

*Figure 7. Imputed $AR_{imp}$, $LAR_{imp}$, $LAR_{imp}$ metrics from (1) - Moody's, (2) - Fitch, (3) - Expert RA(Russia). (A source: Rating agency's data for corporate segment).*

These data are supplied in varying degrees of detail by the Big Three and sovereign rating agencies in the Annual Default Study. To demonstrate the dynamics of annual assessments of $AR_{imp}$, $LAR_{imp}$, $RAR_{imp}$, we took two rating agencies of the Big Three (Moody's and Fitch) and one sovereign (Expert RA, Russia) (see Fig. 7).

The presented assessments show that Moody's is not inferior to Fitch in terms of the imputed Gini index, however, in terms of target preference, Moody's rating "sees" good corporate companies better and distinguishes "bad" worse than Fitch rating, that is, it has a right-hand preference. Fitch's ROC curve is close to symmetrical ($LAR_{imp} \cong RAR_{imp}$). The discrimination power of the sovereign Expert RA is significantly lower than the representatives of Big Three, but according to the estimates of the second-order metrics, Expert RA rating also "sees" the good ones better, having a predominantly right-hand preference. The difference in target preferences of rating mechanisms for different rating agencies complicates the task of comparing their rating scales. Many scientific works are devoted to the latter problem [6], [11], [2], [9], etc.

Validation of the calibration of rating / scoring models requires binary event data as well as the results of the rating of a portfolio of objects, and it is necessary to take into account the statistical uncertainty before drawing conclusions. It is required to calculate AR, LAR, RAR according to the formulas (8),(9) with fixing the statistical error $\sigma_{max}$ (10) of the calculation of AR. The first thing that is important to evaluate is the first-order metrics $|AR - AR_{imp}|$ vs $\sigma_{max}$ and the sign $(LAR - RAR)$ up to $\sigma_{max}$ to understand the preference of the model. If the result of evaluating the first-order metric on the current portfolio $AR_{imp}$ is consistent with the AR estimate on historical binary data, then the analysis of the second-order metrics can be started to answer the question whether the imputed rating result takes into account the target preference of the model, whether there are distortions in the second-order metrics. For this, the dominant LAR or RAR metric calculated on binary data is compared with the corresponding $LAR_{imp}$ and $RAR_{imp}$.

The approach to assessing the internal rating of clients or applicants, the most practical for implementation in a credit institution, is a parametric calibration of the scoring / rating model for the probability of default, which should correspond to the discriminatory power and target preference of the model. A universal approach to model calibration was proposed in [10], where the right-hand and left-hand preference of the scoring model was taken into account.

## 6. Conclusion

It is presents new specialized metrics for the accuracy of second-order scoring models, which show the target preference of the models - whether it is better to identify "good" rating objects (borrowers) or better to identify "bad" ones, provided that the predictive power of the model being unchanged and determined by the generally accepted first-order metric - the Gini index (AR). There are two second-order metrics, left-hand LAR and right-hand RAR. Their definition is given through an integral representation assuming the ROC curve is continuous, but then two groups of representations of finite numerical formulas for metrics are given. The first of which is based on binary events, the second on the default probability given by the model. Comparison of the results of calculating the metrics allows you to validate the calibration settings of the scoring / rating model and reveals its distortions. If the data for first approach is will be difficult to get, then the second will give an understanding of the target preference of the model.

Examples of calculations for three rating agencies are given. The results indicate that Moody's and Expert-RA have a predominantly right-hand preference, while Fitch has a near to neutral preference in own rating. Using the example of the widespread logit formula for the calibration of the scoring model, we made sure that such a calibration is applicable only for the scoring model of near to right-hand or no more than neutral preference ($LAR \approx RAR$). For scoring models that are have the left-hand preference or strictly the righ-hand one the logit calibration can give significant distortion.

The paper also gives a metrization of the ROC-curve, which is unambiguous and alternative to AR/LAR/RAR, but has an obvious geometric meaning in the trapezoidal representation of the ROC-curve. Metrization $AR/\mu_L,/\mu_R$ allows you to consistently divide the ROC-curve into three areas with high, indifferent and low dependence on the X coordinate, to indicate the area where the discrimination model is quasi-neutral.